\documentclass[final,authoryear,5p,times,twocolumn]{elsarticle}
\pdfoutput=1

\usepackage{graphicx}
\usepackage{amssymb}
\usepackage[pdftex,pdfpagemode={UseOutlines},bookmarks,bookmarksopen,colorlinks,linkcolor={blue},citecolor={green},urlcolor={red}]{hyperref}
\usepackage{hypernat}
\usepackage{upquote}
\usepackage{upgreek}

\journal{Astronomy \& Computing}

\newcommand{\qjras}{QJRAS}
\newcommand{\apj}{ApJ}
\newcommand{\aap}{A\&A}
\newcommand{\aspconf}{ASP Conf.\ Ser.}

\newcommand{\datalink}{\texttt{DataLink}}
\newcommand{\ssap}{\texttt{SSAP}}
\newcommand{\tsap}{\texttt{TSAP}}
\newcommand{\regtap}{\texttt{RegTap}}
\newcommand{\obstap}{\texttt{ObsTap}}
\newcommand{\tap}{\texttt{TAP}}
\newcommand{\obscore}{\texttt{ObsCore}}
\newcommand{\getdata}{\texttt{getData}}
\newcommand{\votable}{VOTable}
\newcommand{\plastic}{\texttt{PLASTIC}}
\newcommand{\samp}{\texttt{SAMP}}
\newcommand{\adql}{ADQL}

\newcommand{\vounits}{\texttt{VOUnits}}

\newcommand{\splat}{\textsf{\small SPLAT}}
\newcommand{\splatvo}{{\textsf{\small{SPLAT-VO}}}}
\newcommand{\topcat}{\textsf{\small TOPCAT}}
\newcommand{\vospec}{\textsf{\small VOSpec}}
\newcommand{\specview}{\textsf{\small Specview}}
\newcommand{\dachs}{\textsf{\small DaCHS}}
\newcommand{\aladin}{\textsf{\small Aladin}}
\newcommand{\gaia}{\textsf{\small GAIA}}
\newcommand{\KAPPA}{\textsf{\small KAPPA}}
\newcommand{\specx}{\textsf{\small SPECX}}
\newcommand{\spefo}{\textsf{\small SPEFO}}
\newcommand{\intep}{\textsf{\small INTEP}}
\newcommand{\pleinpot}{\textsf{\small PLEINPOT}}
\newcommand{\IRAF}{\textsf{\small IRAF}}
\newcommand{\IDL}{\textsf{\small IDL}}
\newcommand{\iris}{\textsf{\small Iris}}
\newcommand{\cassis}{\textsf{\small CASSIS}}
\newcommand{\python}{\textsf{\small Python}}
\newcommand{\Sesame}{\textsf{\small Sesame}}
\newcommand{\Starlink}{\textsf{\small Starlink}}

\newcommand{\almaot}{\textsf{\small ALMA-OT}}
\newcommand{\vodesktop}{\textsf{\small VO-Desktop}}
\newcommand{\beanshell}{\textsf{\small BeanShell}}
\newcommand{\subversion}{\textsf{\small Subversion}}
\newcommand{\git}{\textsf{\small Git}}

\newcommand{\ascl}[1]{\href{http://www.ascl.net/#1}{ascl:#1}}

\begin{document}
\begin{frontmatter}

\title{Spectroscopic Analysis in the Virtual Observatory Environment with SPLAT-VO}

\author[OND]{Petr \v{S}koda\corref{cor1}}
\ead{skoda@sunstel.asu.cas.cz}
\author[DUR]{Peter W. Draper}
\author[HDB]{Margarida Castro Neves}
\author[VSB]{David Andre\v{s}i\v{c}}
\author[COR]{Tim Jenness}
\cortext[cor1]{Corresponding author Tel. +420\,323620361}

\address[OND]{Astronomical Institute of the Academy of Sciences,Fri\v{c}ova~298, 251\,65, Ond\v{r}ejov, Czech Republic}
\address[DUR]{Department of Physics, Institute for Computational
Cosmology, University of Durham, South Road, Durham DH1 3LE, UK}
\address[HDB]{Universit\"a{}t Heidelberg, Astronomisches Rechen-Institut,
M\"o{}nchhofstra\ss{}e 12--14, 69120 Heidelberg, Germany}
\address[VSB]{Department of Computer Science, Faculty of Electrical
Engineering and Computer Science, V\v{S}B --- Technical University of Ostrava\\
 17. listopadu 15, 708 33 Ostrava-Poruba, Czech Republic}
\address[COR]{Department of Astronomy, Cornell University, Ithaca, NY 14853, USA}

\begin{abstract}
  \splatvo\ is a powerful graphical tool for displaying, comparing, modifying
  and analyzing astronomical spectra, as well as searching and retrieving
  spectra from services around the world using Virtual Observatory (VO)
  protocols and services. The development of \splatvo\ started in 1999, as part
  of the \Starlink\ StarJava initiative, sometime before that of the VO, so
  initial support for the VO was necessarily added once VO standards and
  services became available. Further developments were supported by
  the Joint Astronomy Centre,
  Hawaii until 2009. Since end of 2011 development of \splatvo\ has been
  continued by the German Astrophysical Virtual Observatory, and the
  Astronomical Institute of the Academy of Sciences of the Czech Republic.
  From this time several new features have been added, including support for
  the latest VO protocols, along with new visualization and spectra storing
  capabilities. This paper presents the history of \splatvo, it's
  capabilities, recent additions and future plans, as well as a discussion on
  the motivations and lessons learned up to now.
\end{abstract}
\end{frontmatter}

\section{Introduction}
There are many tools for the analysis of astronomical spectra.  The most
commonly used of these are the various packages of MIDAS\
\citep[][\ascl{1302.017}]{1992ASPC...25..115W}, \IRAF\
\citep[][\ascl{9911.002}]{2012ASPC..461..595F}, and \Starlink\
\citep[][\ascl{1110.012}]{1982QJRAS..23..485D}, together with scripts written
in \python\ \citep[e.g.,][]{2013A&A...558A..33A} and \IDL\
\citep[e.g.,][]{1993ASPC...52..246L}.  Very advanced processing is still often
done in custom (especially FORTRAN) programs used by small communities.  These
tools tend to work with specially formatted files (quite often ASCII tables)
stored on local disks.

The common tasks for astronomers during spectral analysis consist of finding
the required spectra in various archives, downloading them, understanding and
extracting metadata (such as date of exposure, spectral range, and flux units),
converting to the custom format used by the particular tool, visualizing
selected spectra and  removing the bad ones (for example if they are noisy or
have excessive numbers of spikes). They can finally run some massive processing
or analysis on an improved list.  Very often the measurement of some parameter,
such as the radial velocity or equivalent widths of selected lines, is
accomplished interactively and the spectra must be adjusted by zooming, panning
and scaling.

A real scientific analysis of thousands of spectra is a very tedious task even
from the same instrument, but the real modern challenge is an analysis of
multi-wavelength spectra from different instruments, where all the metadata are
different, there are various flux and wavelength, frequency or energy units
and where the formats of storage also vary (including simple FITS, binary
tables, and ASCII lists).

Making the handling of spectra obtained from world-wide distributed
heterogeneous archives simple and efficient was the main motivation for
developing the International Virtual Observatory Alliance Simple Spectral
Access Protocol \citep[IVOA \ssap;][]{ssap}. The key elements of this
\texttt{HTTP}-based protocol are a standard spectral server discovery mechanism
(a Google for astronomical data) called IVOA Registry \citep{registry} and  the
standard \votable\ data format \citep{2004tivo.conf..118O}, that is used to
describe the metadata of discovered spectra in a limited vocabulary of
obligatory parameters.

\subsection{The SSA protocol}
\ssap\ allows astronomers to find all the world-wide VO-compatible archival
spectra of their favourite objects, and it can also be used to get spectra for
all the objects in a `circular' region of interest on the sky. These simple
queries can be refined using spectral ranges, dates of observation as well as
more subtle criteria like spectral resolution or target class (even spectral
type) if the server supports corresponding optional parameters. More capable
servers provide extended facilities for converting spectra on-the-fly to the
required format, and can create previews of spectra in \texttt{PNG} or deliver
the fluxes and wavelengths in ASCII or CSV tables.

\subsection{SSAP-compatible visualization clients}
Although processing large amounts of data is more convenient when using \ssap\
from a scripting language for batch processing, a crucial part of understanding
data is only gained by using advanced visualization and interactive
exploration.  Therefore, a very important part of the VO infrastructure is
represented by VO-compatible spectra visualization clients. There are currently
several tools for interactively manipulating spectra in the VO.

\vospec\ \citep[][\ascl{1205.011}]{2005ASPC..347..198O}, developed by ESA, is
designed strictly on principles of IVOA standards and was rather focused on
the analysis of high energy and space-based spectra with absolute flux
calibration, lacking support for some common functions and custom data formats
used in ground-based stellar spectroscopy. The most critical 1D FITS image
format was added just recently. \vospec\ contains quite complex non-linear
model-fitting package and can handle photometric points as well as Spectral
Energy Distributions (SEDs).

The other clients were developed as general spectral analysis tools prior to
the conception of the VO, and have subsequently been extended to add support
for querying and downloading spectra using \ssap.  The advantage of this
approach is the support for many legacy data formats and various
domain-specific functionality, better tuned to requirements of scientists not
familiar with VO technology.

\specview\ \citep[][\ascl{1210.016}]{2002SPIE.4847..410B}, developed at STScI,
was intended mainly for handling spectra from \emph{Hubble Space Telescope}
instrumentation and some other NASA missions, but gradually transformed into a
VO-compatible general client supplemented with a library of theoretical spectra
and various models for SED generation.  While \specview\ is a very powerful
tool compatible with a lot of very special formats from \emph{HST}, \emph{IUE},
\emph{FUSE}, \emph{ISO} and \emph{GALEX} satellites, allowing for the advanced
comparison of observations with a wide range of theoretical spectral models
(such as power laws, Bremsstrahlung, emission line profiles, interstellar
extinction, dusty rings or whole Kurucz library of spectral types), it is less
flexible in the support of various \ssap\ features and is falling behind the
rapid evolution in IVOA standards.

Because of its powerful spectral model handling \specview\   was  selected as a
core module for the visualization of spectra and SEDs in the VAO Spectral
Energy Distribution Analysis Tool \iris\
\citep[][\ascl{1205.007}]{2014ASPC..485...19L}.

The specific domain of \cassis\ \citep[][\ascl{1402.013}]{2011IAUS..280P.120C}
is the visualization and analysis of  sub-millimetre  spectra (with special
emphasis on data from the \emph{Herschel Space Observatory} as a plugin for its
HIPE pipeline \citep[][\ascl{1111.001}]{2012ASPC..461..733B}) using various
databases of atomic and molecular transitions and several physical models of
the interstellar environment. In order to be able to directly use spectra from
VO resources, an \ssap\ interface was added recently.

The final VO-compatible client for spectra analysis and the main subject of
this article is \splatvo\ \citep[][\ascl{1402.008}]{sun243}.  \splatvo\ despite
its early origin was revived recently and is currently under intensive
development, serving as a test-bed for newly proposed VO standards and special
features.

\section{History of SPLAT-VO}
\subsection{SPLAT}
The \splatvo\ project is an extension to an older one simply called
\splat\ \citep[][\ascl{1402.007}]{2002ASPC..281..513B}.  The name \splat\
is based on the words SPectraL Analysis Tool, an application for the
display and analysis of spectral data. The development of \splat\ itself
was started by the \Starlink\ Project \citep{1982QJRAS..23..485D} in
1999 in response to a perceived need within the UK astronomy community
for a spectral domain tool with capabilities similar to those of the
successful \gaia\ tool \citep[][\ascl{1403.024}]{2000ASPC..216..615D}
that supported imaging data. \splat\ was also a leader project for
moving future developments into the Java language, with the obvious
benefits of improved portability, modern language capabilities (OOD,
OOP) and core-level support for features like UIs and Internet
protocols and services. This work eventually led to the formation of
what became the StarJava project, containing also the well-known table processor \topcat\ \citep[][\ascl{1101.010}]{2005ASPC..347...29T}.

\subsection{SPLAT-VO}
The first VO capabilities added to \splat\ where released in 2005
\citep{2005ASPC..347...22D}, as support for querying VO registries for Simple
Spectral Access Protocol servers was added. A selection of these servers could
then be queried for any spectra they contained within a region on the sky and
these could then be selected and displayed. \splat\ already had the ability to
match a wide range of spectral coordinate systems and flux units, so the
spectra could be displayed within the same plots for direct comparison. Since
then \ssap\ support has been increased to support version 1 of the standard and
various enhancements like being able to query for the \ssap\
\texttt{TARGETNAME}, not just a region, have been added.

The second phase of VO developments in \splatvo\ were the incorporation into
the \vodesktop\ \citep{2008ASPC..394..251W}, so that spectra and tables could
be shared between applications.  Initially, in 2006, this used the \plastic\
\citep{2007ASPC..376..511T} protocol and then, more recently, \samp\
\citep{2012ASPC..461..279T}.

In 2011 the development of  \splatvo\ was resumed by the German Astrophysical
Virtual Observatory (GAVO) in Heidelberg, Germany and in  the Astronomical
Institute of the Academy of Sciences of the Czech Republic (AI~ASCR).  Because
of the requirement for a spectral domain application with VO capabilities and
good scientific analysis functions, GAVO  decided to actively support
developments now that \splatvo\ was only being supported by unfunded best
efforts, concentrating on an improved VO interface, benefitting from the
in-house development of the VO server suite \dachs.  Around the same time
requirements for some modifications of scientific functions (with emphasis on
optical stellar spectroscopy) identified during the everyday analysis of
spectra of emission line stars in the Stellar department of AI~ASCR, resulted
in Bachelor and successive Diploma theses focused on the implementation of new
analytical and visualization functions in collaboration with
V\v{S}B---Technical University of Ostrava.

In this way \splatvo\ can continue to be improved, and kept up-to-date with the
changing requirements of both end users and data providers.

\section{SPLAT-VO internals}
\subsection{Data formats}
The \splatvo\ spectral data model is specified by a single internal interface
that is concretely implemented for the various data formats that are
supported. Having a single interface decouples the data model and makes it
practical to support a wide range of formats, but necessarily means that some
format specific features will not be supported and can be lost when making
transformative changes like saving back to file.

The \splatvo\ data model is simple only requiring that spectra have a name,
coordinate system and some data values. Other supported items are data errors,
fuller descriptions, keyed properties (key, value, description), data units
and labels, and the underlying dimensionality (if the spectrum is being
extracted from a 2D or 3D array). Spectra can also have mutable properties,
like the names of the table columns that contain the various data values.

Current implementations give \splatvo\ access to spectra in simple text files,
FITS \citep{2010A&A...524A..42P}, NDF \citep{ndfjenness}, NDX
\citep{2003ASPC..295..221G}, as well as FITS and VO tables. Tables are
supported using the STILTS \citep[][\ascl{1105.001}]{2006ASPC..351..666T}
library. In addition to these spectra can also be memory based so that they can
become modifiable and be derived from data with higher dimensionalities.

\subsection{Internal spectra handling}
The presentation of all data types through a single interface works especially
well when coupled with a further, higher-level, abstraction class. This
abstraction is in effect how all the analysis and visualization code throughout
\splatvo\ see a `spectrum' (these two elements that decouple an abstraction
from the underlying implementations are in fact the well established `Bridge
Pattern').

This use of a single abstraction to all data formats makes it easier to write
consistent code and extend the basic internal spectrum with new facilities
(methods that do things like extraction sections, replace bad values etc.) as
well as associate useful properties like graphics and formatting rendering
hints, data and coordinate range limits. Extending the underlying model is also
easy as new features strictly need only be handled in the abstraction (although
clearly that would not be interesting) and new spectral types can extend the
base abstraction (the basic spectral type is extended to also offer spectra
that render as line identifiers and cached copies of remote spectra).

\subsection{Presentation}
The presentation of spectra is based on the model-view-controller pattern that
pervades Java UIs and the same spectral data can be seen and operated on in
many different guises, that is as a number of line plots, tables, property
lists etc. Memory based spectra are also directly modifiable (as simple edits
to table cells or generally using expressions). Through the
model-view-controller infrastructure all changes are propagated to all views.

\subsection{Coordinate systems and units}
Early in the design of \splat\ it was decided that it would not be possible to
replicate the considerable effort that had already been put into a
comprehensive system for handling coordinate systems and units, including full
support for reading and writing FITS WCS \citep{2006A&A...446..747G}. This led
to the development of a JNI library for the \Starlink\ AST library
\citep[][\ascl{1404.016}]{1998ASPC..145...41W,2012ASPC..461..825B}, JNIAST. AST
itself is written in standard C so compiling it on many platforms is relatively
easy, but, as discussed in Sec.\ \ref{sec:jniast-lesson}, this remains a
portability issue at the heart of \splat\ limiting one of the initial aims.

Some of the advanced features that \splatvo\ offers based on AST are support
for spectral coordinates in wavelength, velocity, frequency and energy. These
can have standards of rest, so transformations of rest frame are also supported
(important for velocity and laboratory data). \splatvo\ also supports double
side-band spectra for sub-millimetre spectra and flux unit matching. Many of
these features have been improved in AST over time and \splatvo\ gains from
this.

AST supports all the units specified in the FITS standard
\citep{2010A&A...524A..42P} and can calculate scaling factors to allow matching
of spectra with different flux units. Adding full \vounits\ support
\citep{vounits} to \splatvo\ would currently require that AST is updated to
match the standard. This would have the added benefit of expanding support for
\vounits\ to the entire \Starlink\ software collection and in particular
enhance the unit support of the VO aspects of \gaia\
\citep{2009ASPC..411..575D}.

\subsection{Scripting of SPLAT-VO}
The various classes of \splatvo\ may be called directly from the \beanshell\
scripting language \citep{niemeyer2013learning}, in principle this allows
complex workflows to be created, and is used to create example scripts that are
part of the standard distribution. These support the batch fitting of single
Gaussian line profiles, as well as composite models consisting of any number of
Gaussian, Lorentzian and Voigt profiles for deblending complex lines (a feature
that has never made it into the full UI). \beanshell\  scripts are also used
for controlling \splatvo\ from command line.

\subsection{The power of SAMP in SPLAT-VO}
Although all \ssap\ clients have powerful capabilities, they are naturally
limited to their built-in features, which will not suit all the requirements of
a wide spectroscopic community. However, almost all VO-compatible tools,
including the important table processor \topcat\ and stellar atlas \aladin\
\citep[][\ascl{1112.019}]{2005ASPC..347..193O}, have built-in support for the
Simple Application Message Protocol (\samp), successor of \plastic. This
supports the bringing together of basic VO applications, like bricks, to build
complex visualization and processing pipelines allowing very complicated
analyses of truly Big Data.  The astronomer may, for example, visualize the
spectra of objects they click on in some imaging survey frame displayed in
\aladin, or identify the visual appearance of galaxies from the spectra which
have been visually selected in \splatvo. \samp\ has successfully been used to
send messages from a web browser displaying previews of a number of spectra to
\splatvo\ where they were subsequently automatically downloaded and then
analyzed.

\section{SPLAT-VO capabilities  for spectra analysis}
\begin{figure*}[t]
\begin{center}
\includegraphics[width=0.8\textwidth]{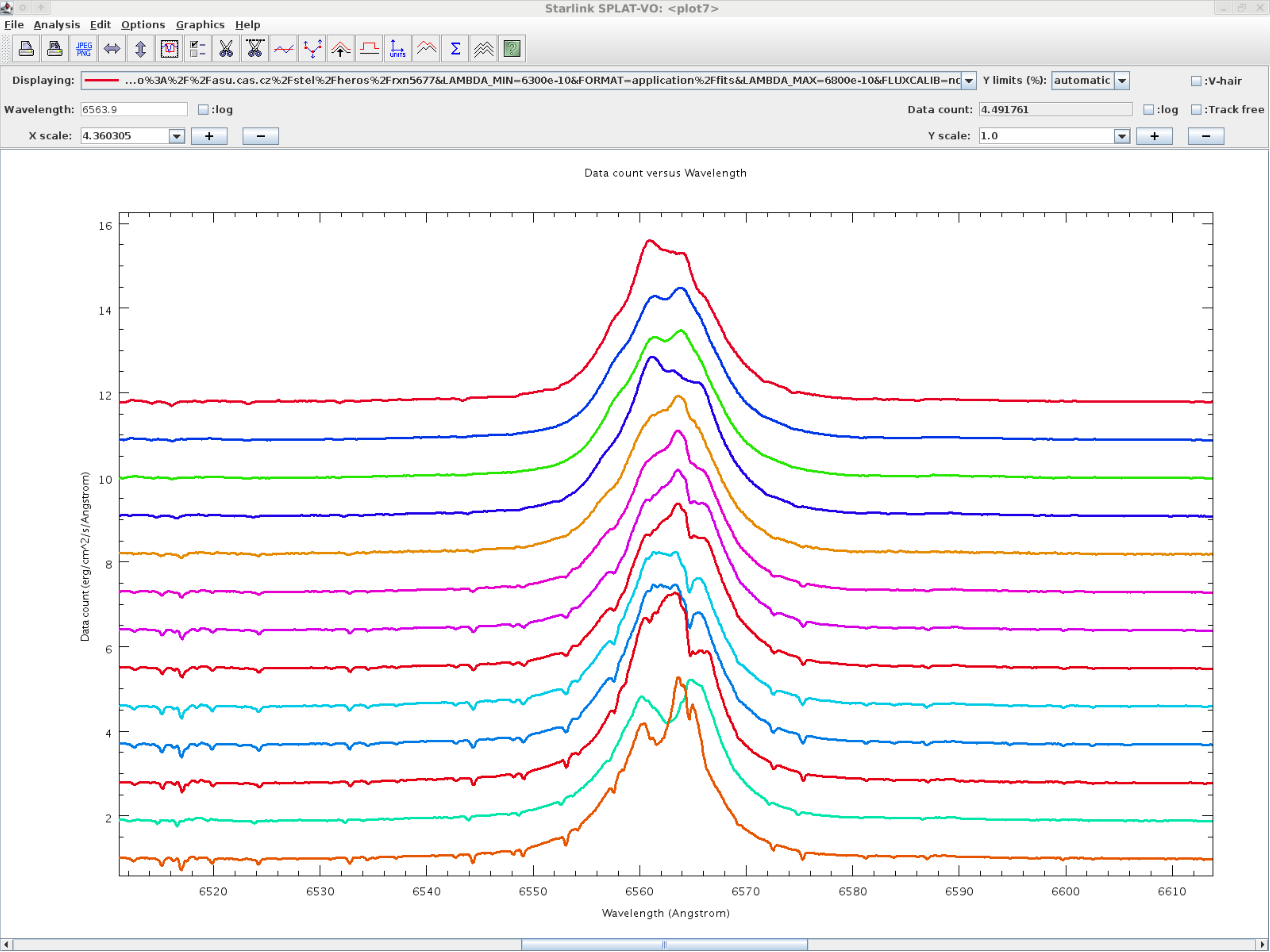}
\caption{The time evolution of the H$_\alpha$ line emission profile in
$\phi$~Per observations obtained with the HEROS spectrograph at Ond\v{r}ejov 2m
telescope in years 2001--2003. The stacking offset of 0.9  makes sure that the changes can be seen easily.}

\label{fig:phiper-heros-stack}
\end{center}
\end{figure*}

\begin{figure*}[t]
\begin{center}
\includegraphics[width=0.8\textwidth]{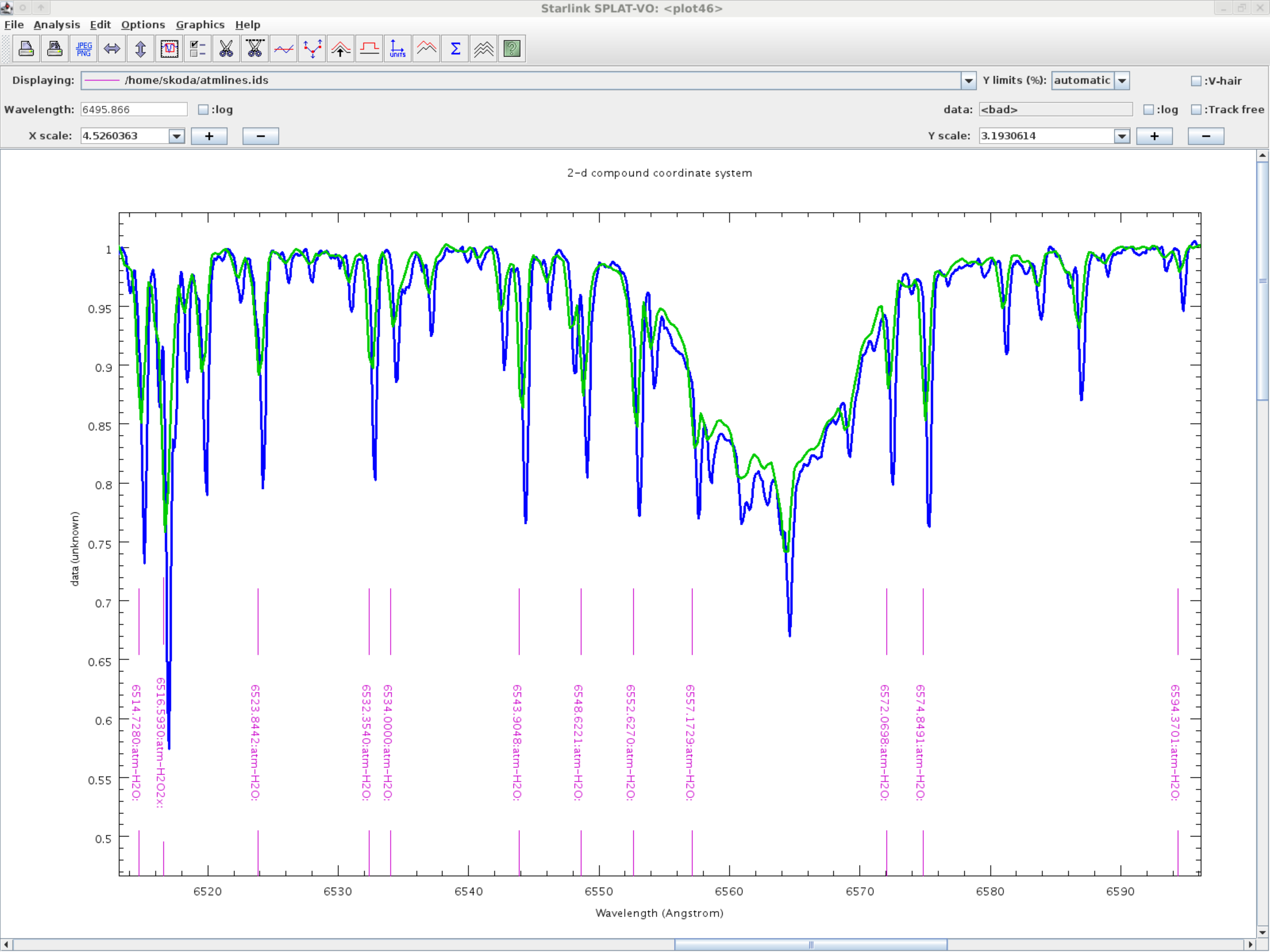}
\caption{Spectrum of $\zeta$~Oph in H$_\alpha$ region  obtained with Ond\v{r}ejov Coud\`e
  spectrograph 700mm camera (green) and HEROS spectrograph (blue). The line
  identifications from custom line list of telluric lines is overplotted in violet. Note the deeper
  telluric lines from HEROS due to its almost double spectral resolving power.}
\label{fig:zetoph2sp-id}
\end{center}
\end{figure*}

\splatvo\ as a general purpose spectral analysis tool supports a very wide
range of capabilities required by astronomers for every-day work outside of the
VO. Input spectra may be loaded directly from local disk files in a number of
formats, including several types of multicolumn ASCII tables, multi-extension
FITS with binary tables or 1D spectrum FITS image, as well as \Starlink\ NDF.

The list of capabilities of \splatvo\ is enormous and are fully described in
the associated documentation \citep[SUN/243;][]{sun243} which is also available
as online help and includes example data from multi-wavelength observations.

Here we give a short overview of features not relevant to VO protocols.
Several features are very specific to the analysis of hot emission line stars.

\begin{itemize}

\item Visual stacking of spectra introducing either constant vertical offsets
or offsets based on functions of values of certain FITS keywords. For example
the line profiles may be ordered by the circular phase computed for a certain
period. Stacking is also useful for visualization of evolution of stellar line
profiles in variable stars.  Fig.~\ref{fig:phiper-heros-stack} shows the
evolution of an emission line profile for the star $\phi$~Per observed with the
HEROS spectrograph \citep{2002PAICz..90....1S} at the Ond\v{r}ejov 2m telescope
during the years 2001--2003.

\item Measurement of radial velocities by the visual matching of a line profile
with its mirrored image based on idea of oscilloscopic comparators.  This is
important for measuring asymmetric and distorted line profiles typical for hot
emission stars. For a detailed explanation see \citet{2007IAUS..240..486P}.
This feature can be operated with a batch-like mode to quickly process a number
of lines and/or spectra using a \emph{visitor list}.

\item Continuum estimation using flexible curves similar to those expected to
be drawn by a human. The \intep\ procedure based on Hermite polynomials
\citep{1982PDAO...16...67H} has been successfully used for decades in fitting
stellar continua (typically of emission-line stars such as Be stars and
symbiotic novae) in the program \spefo\ \citep{1996ASPC..101..187S}. An
overview of the advantages of this procedure is given in
\citet{2008asvo.proc...97S}. Other curves types are also available such as
\citet{Akima:1970:NMI:321607.321609}.

\item Powerful transformation of the spectral axis with respect to various
coordinate systems; air and vacuum wavelength, optical and radio velocities,
frequency, redshift, energy. The transformations can also include various
standards of rest; topocentric, heliocentric, dynamic and kinematic local.

\item Built-in spreadsheet processor supported by special astronomical
functions. Allows the modification/transformation of coordinates and data
values. Astronomical functions include the spectral line profile models.

\item Filtering with smoothing (mean, median, rebin) and denoising with
wavelets.

\item Line fitting using Gaussian, Voigt or Lorentz profiles.

\item Statistics of selected regions, mean, median, sum, mode, variance, skew,
kurtosis, rms, quantiles.

\item Region removal, replacement and extraction. Also replacement of part of
spectra with an interpolated curve.

\item Comprehensive overlay graphics toolkit with annotations and many
configurable options for the presentation of data. Export of publication
quality output into \texttt{PNG}, \texttt{GIF}, and \texttt{EPS} formats.

\item Animation of spectra - convenient for showing line evolution or
non-radial pulsations.

\item Identification of spectral lines using simple supplied ASCII line lists
and built-in sets of common atomic and molecular lines (includes molecules in
sub-millimetre region). Fig.~\ref{fig:zetoph2sp-id} shows the identification of
telluric lines on a combination of spectra of $\zeta$~Oph obtained with two
different spectrographs of the Ond\v{r}ejov 2m telescope.

\item Simple operations on spectra such as addition, subtraction, division and
continuum fitting.

\end{itemize}

\section{SPLAT-VO SSAP interface}
The \ssap\ interface in \splatvo\ is clearly the most important tool for
working with the VO and remains under active development. The task of the
interface is to help the user construct a valid \ssap\ query by filling out a
simple form of parameters, and then arrange for that query to be sent to a list
of selected \ssap\ servers and finally make the query response available so
that any spectra can be selected for download.

Queries can involve supplying the coordinates (\texttt{POS}) and size
(\texttt{SIZE}) defining a 'circular' region on the sky (a so called `cone
search'), or just the name of an object and size, the name being resolved into
a position by the CDS name resolver \Sesame . Other form elements provide
refinement of the query to include things like spectral ranges (\texttt{BAND}),
the date and time of observation (\texttt{TIME}), status of the wavelength
calibration (\texttt{WAVECALIB}), flux calibration (\texttt{FLUXCALIB}) and
data format (\texttt{FORMAT}).  In fact recent additions to \splatvo\ now offer
control of all the optional parameters supported by a selected \ssap\ server.

Once spectra have been selected they are downloaded from the \ssap\ server
(using the \texttt{URL} \texttt{accref} returned in the \ssap\ response) they
can be worked on in the same fashion as any locally available spectra and saved
to local disk etc.

The list of \ssap\ servers may be updated using one of several VO registries.
Local or development services not yet registered may be added manually. The
server list may be edited so that the query is only sent to collections of
servers with say similar data or can be restricted to only one service that may
be queried for detailed specific operations.

\begin{figure*}[th]
\begin{center}
\includegraphics[width=0.8\textwidth]{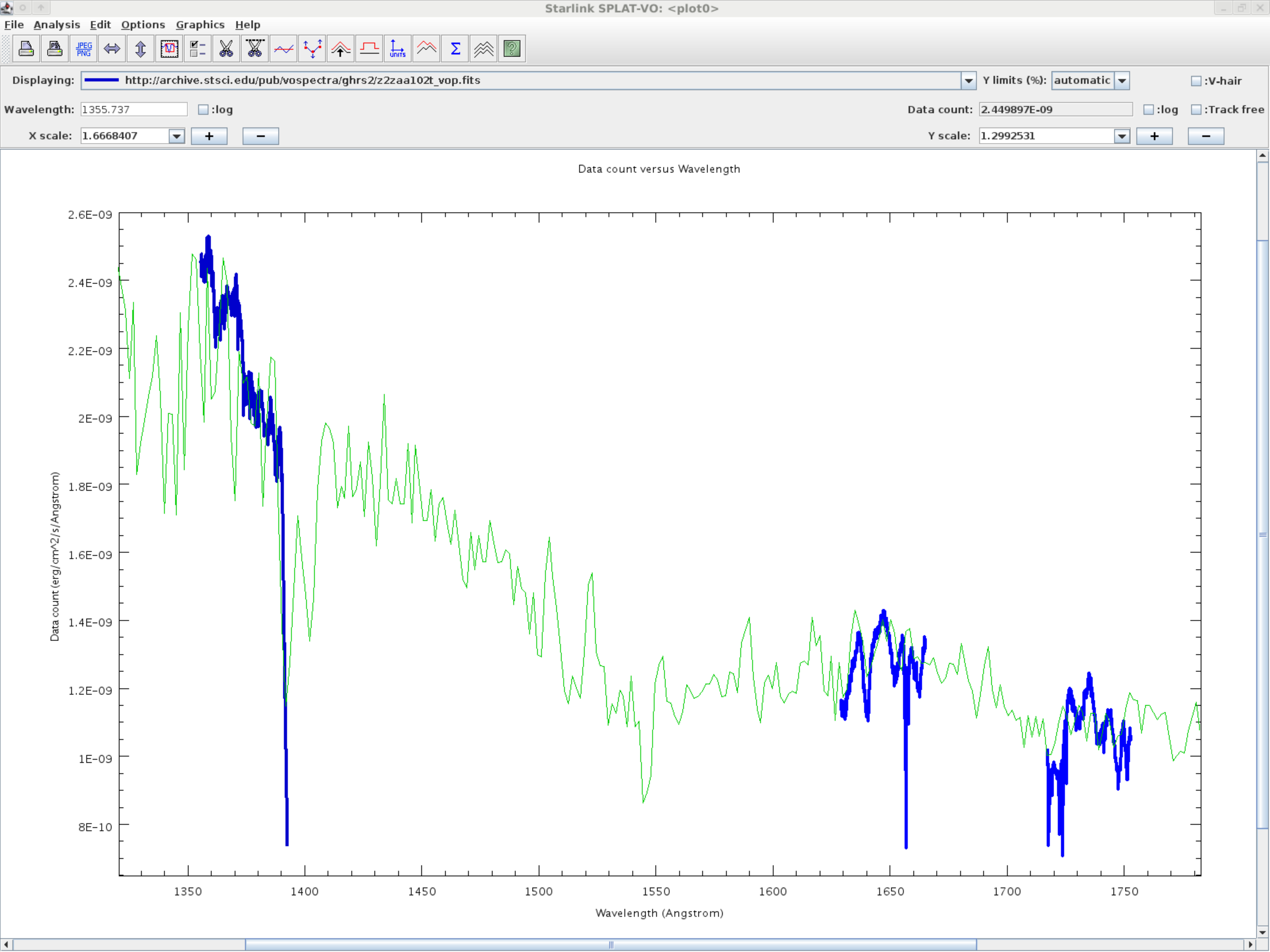}
\caption{Composite plot of UV spectra of $\phi$~Per from the \emph{HST} GHRS
spectrograph (thick blue short pieces) together with \emph{IUE} low resolution
spectra from SWP camera, (thin green line).}
\label{fig:iuehst2}
\end{center}
\end{figure*}

The following example shows the absolutely calibrated spectra of the emission
star $\phi$~Per. We have selected the \emph{Hubble Space Telescope} with the
Goddard High Resolution Spectrograph and the \emph{IUE} low resolution spectra.
The composite plot of selected  UV spectra is in Fig.~\ref{fig:iuehst2}.

\subsection{ Theoretical spectra access}
\begin{figure*}[t]
\begin{center}
\includegraphics[width=0.8\textwidth]{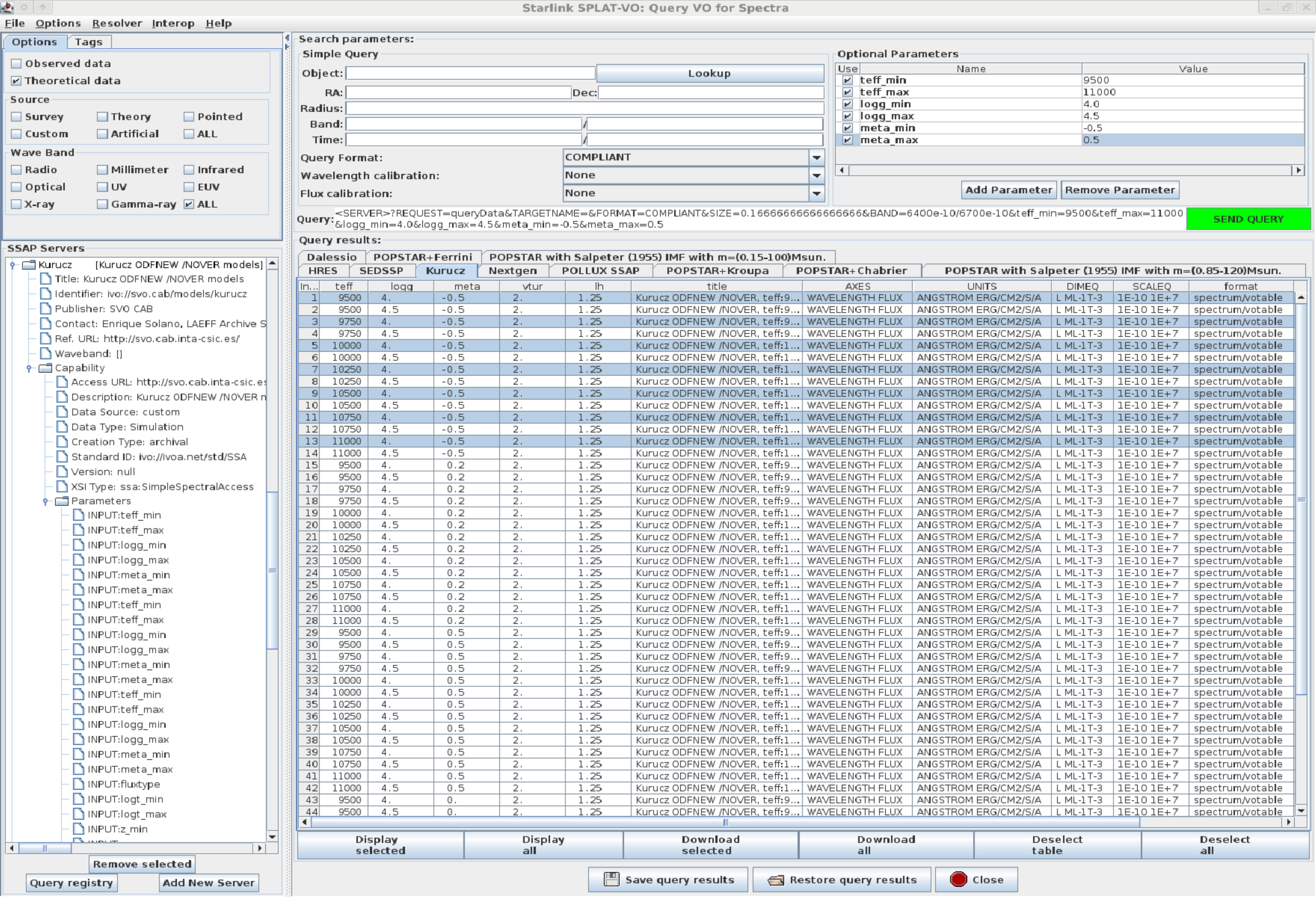}
\caption{\tsap\ Query for Vega-like Kurucz models. Note the additional
parameters.}
\label{fig:TSAP-query}
\end{center}
\end{figure*}

\begin{figure*}[t]
\begin{center}
\includegraphics[width=0.8\textwidth]{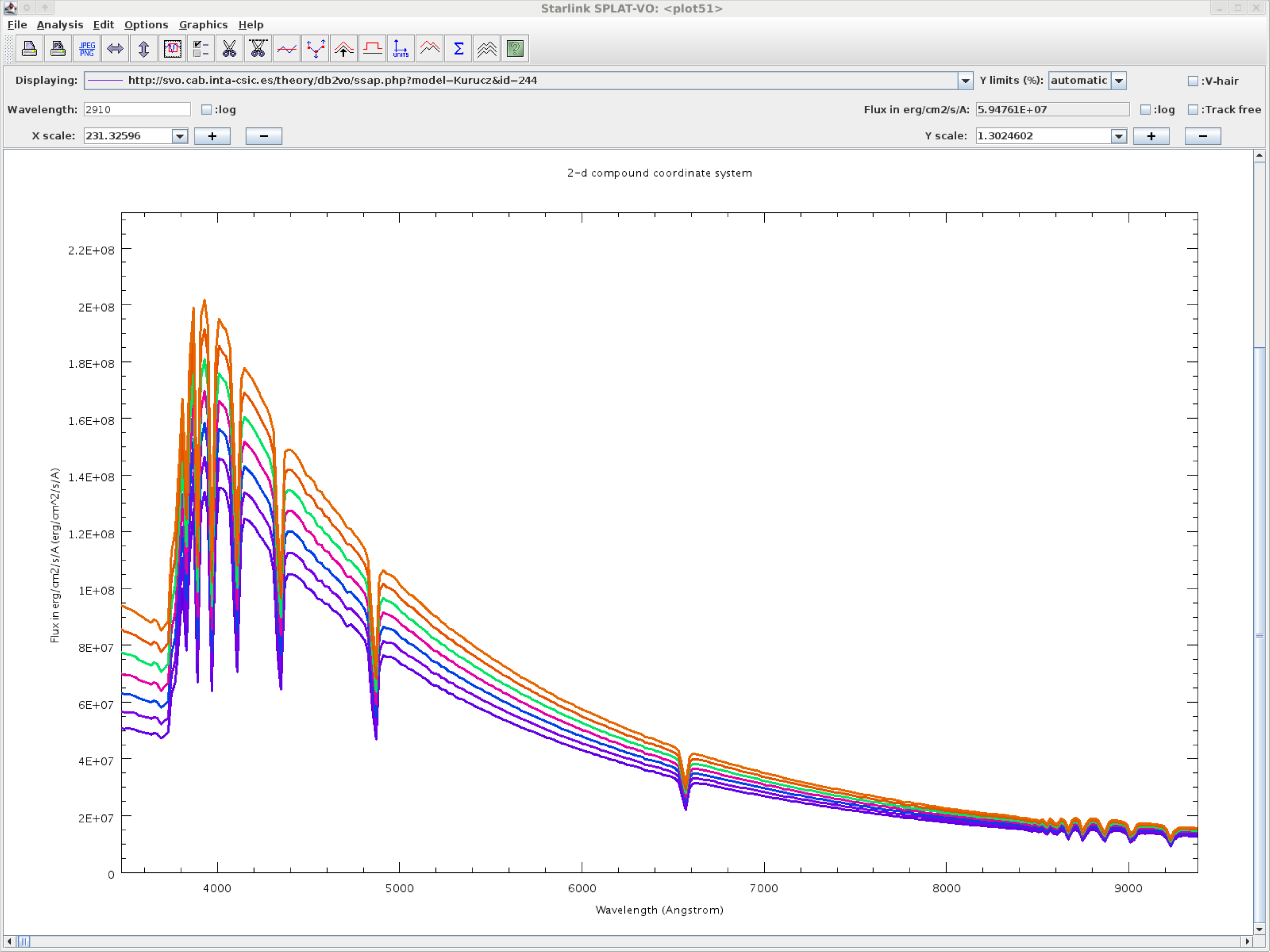} \caption{Zoomed plot of
Vega-like Kurucz models manually selected in the previous
\tsap\ Query window (see Fig.~\ref{fig:TSAP-query}). }
\label{fig:TSAP-plot}
\end{center}
\end{figure*}

In a similar way to observational spectra, synthetic spectra can also be
obtained from libraries available in the VO (e.g., Kurucz models, Rauch's
non-LTE model spectra, TLUSTY hot stars, Salpeter, Dusty). They are selected by
a radio button, which changes the role of parameters used in the query. The
extended protocol called Theory-\ssap\ \citep[hereafter TSAP;][]{ssap} uses
various physical parameters like $T_{\rm eff}$, $\log g$, metallicity etc. for
queries instead of position, wavelength, region and date/time range.
Unfortunately obligatory metadata do not exist for all required physical
parameters, so it is necessary to just try out what is available on a given
server.  It is then necessary to individually select suitable spectra from a
multidimensional table combining all the parameter ranges required.
Fig.~\ref{fig:TSAP-query} and Fig.~\ref{fig:TSAP-plot} show the query and
result of selecting several Kurucz models for Vega.

\section{Recent additions}
\subsection{Motivation}
As the VO evolves, adding new protocols and data models, \splatvo\ needs to
evolve accordingly.  \ssap, as the name says, is a simple protocol, and has
some limitations. Some features like cut-outs or flux calibration, which are
often needed, have to be performed in a second work step after downloading the
spectra. It is much more efficient to have it done on-the-fly at the server and
to download the spectra exactly as needed. To overcome the limitations of
\ssap, the new \datalink\ protocol \citep{datalink} was used. \splatvo\ is one
of the first client implementations of \datalink, in this way also contributing
to its testing and development from a client point of view.

Besides \ssap, spectra can also be retrieved using the \obscore\ data model
through Table Access Protocol  \citep[known as \obstap;][]{obstap}. Its
implementation in \splatvo\ allows spectra to be retrieved also from \obscore\
services, and the \obstap\ \adql\ \citep{adql} search offers a more powerful
search mechanism as well. The implementation of new VO protocols like \obstap\
and \datalink\ have been done side-by-side with their server-side
implementation in \dachs\ \citep[Data Center Helper Suite;][]{dachs}.

\splatvo's user interface has been improved. An improved server selection
interface and \ssap\ search by metadata parameters are some of the new features
added, which are listed in the next subsections.

\subsection{New features in VO access}
\subsubsection{SSAP service selection}
In the earlier versions of \splatvo\ a list of services was presented, and the
user could select the services to be queried, as well as remove uninteresting
services from the list. Information about the services of interest had to be
retrieved from somewhere else, as the service metadata was not used.  This
simple server selection interface has been improved by a more detailed
selection, users can choose a set of services that may contain data of
interest, based on the service's metadata taken from the registry (data source,
wave-band).

There are two ways of selecting services. The first way is to choose according
to the data source (such as a survey, pointed observation, or simulation) or
wave-band.  For example, the user wants to select services containing pointed
observation sources in optical wavelength, so this can be selected in the
interface and only the services which contain this information in the metadata
will be selected.  The other way is useful in the case when users know exactly
the services they want to query, so they may create an own tag containing the
chosen services, and give it a name, such as ``My Favourite Services'', which
will be saved and loaded the next time \splatvo\ starts.

This implementation relies on the metadata information from the registry
services, which should be correct and complete. Unfortunately many services
information from the registries are not complete, or not correct, what we hope
will be fixed in the future.

\subsubsection{Manual addition of SSAP services}
\splatvo\ gets its list of \ssap\ services by querying a registry service.  A
new function has been added to allow inclusion of a new service that is not
(yet) registered. This is useful to access a local, not public service, or to
test a service before it's included in the registry.

\subsubsection{Metadata parameters query}
In early versions, \splatvo\ contained just a simple cone search using the
\ssap\ obligatory parameters (RA, DEC, band, time, data format) and later added
wavelength and flux calibration state, although much other metadata is
available on servers.  Now \splatvo\ also retrieves the metadata parameters for
each server, and a new user interface allows users to chose metadata parameters
and their values to perform more detailed queries.  This is especially
interesting when querying theoretical spectra by \tsap, which contain many
additional query parameters. Figure~\ref{fig:TSAP-query} shows the use of
additional metadata parameters.

\subsubsection{Simple HTTP authentication}
Simple HTTP authentication can be used to restrict access to data when it is
published to the VO before it should be publicly available, so the owner and a
limited group of users can access it before disclosure. The authentication can
be done on a service basis, where the whole service needs authentication to be
queried, or on a spectrum basis, where only some of the spectra from a certain
service require authentication to be downloaded. In any case, an
username/password window will appear when needed.

\subsubsection{GetData and DataLink}
To overcome \ssap\ limitations and allow server-side processing of spectra like
cut-outs, or format conversions, a protocol called \getdata\ \citep{getData}
has been developed. After successful implementation and testing on \splatvo,
the work on \getdata\ has been stopped. It has been decided to implement these
features using the newer \datalink\ protocol, which will probably be made
available on several services in the near future, and can also be used in
different cases.  \datalink\ resources can provide links with several different
ways to access the data, covering \getdata\ functionality and more.  When
parsing the service response from an \ssap\ query, which is in form of a
\votable, \splatvo\ looks for a \texttt{RESOURCE} element of type
\texttt{service}.  An example is shown below:

{\tiny
\begin{minipage}{\textwidth}
\begin{verbatim}
<RESOURCE ID="aeoadowdpudn" type="service">
  <GROUP name="input">
    <PARAM arraysize="*" datatype="char" name="ID"
    ref="ssa_pubDID" ucd="meta.id;meta.main" value="">
      <DESCRIPTION> The publisher DID of the dataset of interest
      </DESCRIPTION>
    </PARAM>
    <PARAM arraysize="*" datatype="char" name="FLUXCALIB"
    ucd="phot.calib" utype="ssa:Char.FluxAxis.Calibration" value="">
      <DESCRIPTION>Recalibrate the spectrum.  Right now, the only
      recalibration supported is max(flux)=1  ('RELATIVE').</DESCRIPTION>
      <VALUES>
        <OPTION name="RELATIVE" value="RELATIVE"/>
        <OPTION name="UNCALIBRATED" value="UNCALIBRATED"/>
      </VALUES>
    </PARAM>
    <PARAM ID="aemoogtm" datatype="float" name="LAMBDA_MIN"
    ucd="par.min;em.wl" unit="m" value="">
      <DESCRIPTION>Spectral cutout interval, lower limit</DESCRIPTION>
      <VALUES>
        <MIN value="3.3696e-07"></MIN>
        <MAX value="8.7665e-07"></MAX>
      </VALUES>
    </PARAM>
    <PARAM ID="appadowdpudn" datatype="float" name="LAMBDA_MAX"
    ucd="par.max;em.wl" unit="m" value="">
      <DESCRIPTION>Spectral cutout interval, upper limit</DESCRIPTION>
      <VALUES>
        <MIN value="3.3696e-07"></MIN>
        <MAX value="8.7665e-07"></MAX>
      </VALUES>
    </PARAM>
  </GROUP>
  <PARAM arraysize="*" datatype="char" name="accessURL" ucd="meta.ref.url"
  value="http://dc.zah.uni-heidelberg.de/flashheros/q/sdl/dlget"/>
</RESOURCE>
\end{verbatim}

\end{minipage}
}

The \texttt{GROUP} element with name \texttt{input} contains the input
parameters that will be returned to the server.  It must contain an \texttt{ID}
parameter, which refers to the metadata that identifying the required spectra
(normally the \texttt{pubDID} parameter).  The other parameters in this group
are the ones for which the users can set values that will be sent to the
service as a request. The \texttt{accessURL} parameter defines the service URL
to which this request has to be sent.

In the \ssap\ response to a query, the services supporting \datalink\ will be
marked with the  ``\ding{34}''  icon. When one of these services is selected,
the user can activate the \datalink\ feature by clicking on the button with
same name. When activated, a window with a form to enter values to the
\datalink\ parameters will appear. While this window is activated, every
spectrum selected and downloaded from this service will be processed according
to the chosen parameters. When de-selected, the spectra will be downloaded as
they are in the \ssap\ service, without processing.

In the case described above, the \datalink\ resource is in the query response
from the \ssap\ service. In another case, after a query, some services return a
list of spectra pointing not to the spectral data to be retrieved, but to a
\votable\ containing only \datalink\ information with the links to the URLs of
the data. This \datalink\ resource has then to be parsed by \splatvo\ in order
to retrieve the selected spectrum.

\subsubsection{ObsTAP}
In addition to \ssap, spectra can be retrieved using \obstap. This service uses
the Table Access Protocol (\tap) to query metadata from the Observation Data
Model Core Components (\obscore). Currently there are few services implementing
it, and the \splatvo\ implementation is ongoing.  \obscore\ provide standard
metadata attributes that can be used, through the \adql\ query language, to
perform more extensive and detailed queries than in the case of \ssap\ data
discovery.

In the current \splatvo\ implementation, users can select \obscore\ in the main
\splatvo\ window. The \obscore\ browser window will appear, which is in part
similar to the \ssap\ browser window. The user can chose either a similar cone
search interface like the one in \ssap, or a simple \adql\ interface where some
parameters can be set. For more advanced queries, the user can directly edit
the \adql\ expression. After sending the query the results tables will be
displayed, which are in functionality similar to the \ssap\ browser window.

\subsubsection{Visualizing light curves using SSAP}
\begin{figure*}[tbh]
\begin{center}
\includegraphics[width=0.8\textwidth]{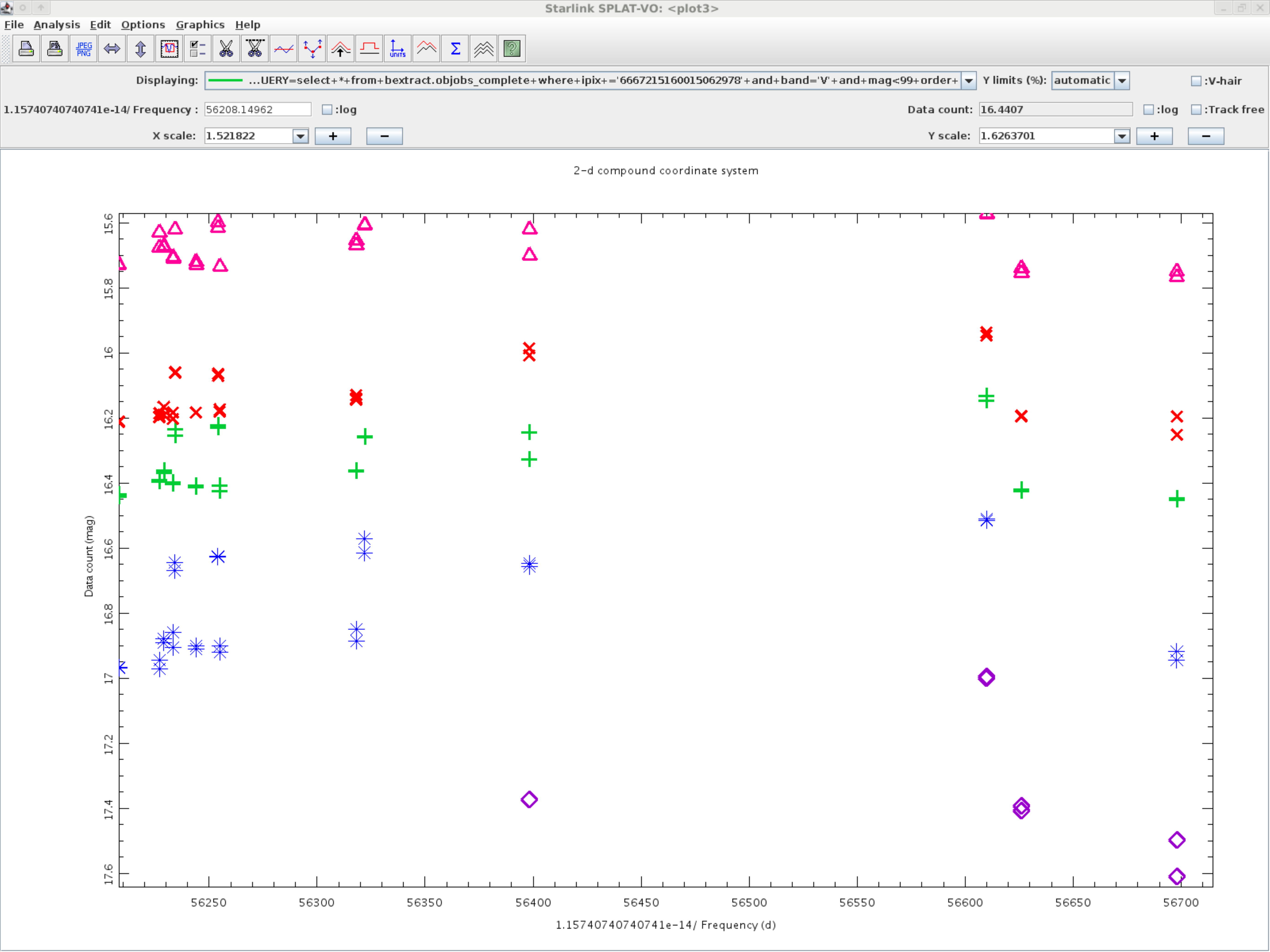}
\caption{Multicolour light curves of Cepheid OGLE-SC7-127550 from OSPS. The
manual modification of default line types into various markers and flip of
vertical axis was applied on data obtained by an experimental \ssap\ light
curve protocol.  The measurements in Johnson-Cousins filters correspond to U
(diamonds), B (asterisks), V (pluses) R (crosses) and I (triangles).  }
\label{fig:OGLE-SC7-127550_plot}
\end{center}
\end{figure*}

As a proper standard for displaying photometric light curves or time series of
other variables is not available from the IVOA, there have been several
attempts to exploit the visual similarity of spectra and light curves
pretending the spectral axis is the temporal one.  So the \ssap\ was
successfully used as a transport protocol for delivering light curves in binary
table format. In principle the table may contain many columns displayed as
dependent variable against the various time axes with interactive menu.  The
current limitation only requires the time axis expressed as floating point
variables, which limits the possibility of using strings e.g., dates, time
stamps or even the names of CCD images used to extract given photometric point
on light curve.  An example in Fig.~\ref{fig:OGLE-SC7-127550_plot} shows the
light curves obtained from the Ond\v{r}ejov Southern Photometry Survey
\citep{skoda_adassxxiii} ongoing at the Danish 1.5m telescope DK154.

\subsection{New features in spectral analysis}
\label{davids_functions} To improve the analytic capabilities and general
usability the following features were added to the non-VO capabilities of
\splatvo\ \citep{and146bcthesis}:

\subsubsection{Saving spectra stack in multi-extension FITS format}
Previous versions of \splatvo\ were able to save the global list of spectra to
a binary format, which was in fact a dump of the native serialization of
spectra objects. Sometimes, there may be a need to save that list to a more
universal, standardized format. For example, \splatvo\ has many operations on
spectra, which result in a new spectrum that is added to a global list of
spectra (e.g., cutting a part of spectrum). Saving the list of such spectra to
a standardized format would allow their use in other tools offering
capabilities that \splatvo\ does not support.  This may be helpful for the
preparation of large lists of normalized line profiles for Fourier
disentangling or asteroseismology studies.

\splatvo\ now allows the user to save the global list of spectra to a much more
universal FITS format, where every spectrum  is represented by its own
(\texttt{IMAGE}) extension.  The only disadvantage compared to the native dump
format is that settings related to the way that the spectra are rendered are no
longer available as there is no standard for metadata of presentation styles
and graphics attributes.

\subsubsection{Visual spectrum selection and highlighting}
In earlier versions of \splatvo, when working with multiple spectra displayed
in one common plot, it was difficult to work out which spectrum corresponded to
one in the global list, so that it could be worked on independently.  Also the
selection of, for example, a noisy spectrum and its removal from the plot
window could be done only by a trial-and-error procedure and therefore was
quite problematic and time-wasting.

This has been improved and now when a spectrum is selected in the global list,
\splatvo\ will highlight it in all the plot windows that it is displayed in.
The highlighting has the form of a few-seconds blinking in an inverted colour
and is performed sequentially (highlighting in one plot window comes after the
previous one is finished), which gives to the user a perfect sequential
overview of the spectrum's plots.

\subsubsection{Cut window: saving the ranges}
In the cut regions tool, it is possible to read the ranges of regions from a
local text file. But what if a visual selection of ranges from a currently
plotted spectrum is to be saved?  \splatvo\ can now save the currently
defined ranges to a local file in the same format as used for reading.

\subsubsection{Cut window: working with multiple spectra}
Previously any regions defined by the cut regions tool only operated on one
spectrum (known as the current spectrum). This tool now has a table of all
currently plotted spectra which can now be selected so that any region
operations are applied to them all.

\subsubsection{Display of spectra received by SAMP in one window}
In previous versions of \splatvo, all spectra received via \samp\ messages were
opened in their own plot windows. Since it would be useful to also display
spectra in one plot to make comparisons, a new control to toggle this behaviour
on and off has been added.

\section{Lessons learned}
During the development of \splatvo\ a number of lessons have been learned that
may affect future developments of similar tools. In this section we summarize
these issues.

\subsection{Using JNI versus ``pure'' Java}
\label{sec:jniast-lesson}
The question should probably be asked if using a JNI wrapper to the AST library
was a good choice or not. As with most things, we believe this was a good idea
and also bad. The judgement of how bad depends on how the ultimate cost benefit
analysis works in the longer term, but either way it is difficult to deny that
a pure Java solution would have been preferable.

The positives of this choice are that it made the development of \splatvo\
possible on useful time-scales as we were able to immediately benefit from all
the work that had gone into the AST library, as AST is very much more than just
a system for handling coordinates and units. It also insulates against the
often messy need to parse FITS headers and offers the ability to transform
easily between all compatible coordinates and units. Even today some 15 years
later, AST is still believed to be a class leader in these areas.

Avoiding this work was also pragmatic as a re-write into Java would have been a
lot of work that the effort just wasn't available for and still isn't. In the
intervening years there have also been new features added to AST, such as the
dual-sideband support, which was vital for sub-millimetre work, and the
adoption of \splat\ by the Joint Astronomy Centre (JAC). The development of
\splat\ has also been a useful testing ground for the development of AST itself
(like the introduction of thread safety, so not just the handling of spectral
coordinates), closing a virtuous circle.

The downsides are only the obvious ones, not having a pure Java solution
compromises the portability, so we need to make an additional effort for all
the supported platforms. There is also the non-trivial time writing the JNI
layer itself takes. Writing JNI interface code is more challenging than those
more commonly found for scripting languages.

\subsection{Open sourcing from the beginning}
The decision was made very early on for the \splat\ source code to be made
publicly available, first via CVS on a \Starlink\ server, then via \subversion\
on a Joint Astronomy Centre server and finally using \git\ on Github. We feel
that this openness contributed directly to the software being picked up by the
GAVO project, saving the application from stagnation once the direct funding
for it was dropped.

\subsection{Local spectral line catalogues are not sufficient}
Currently \splatvo\ uses a local text file containing molecular transitions
that are thought to be of interest and allows the user to provide external
lines via a text file of a similar format. The internal list is dominated by
sub-millimetre lines required by the JCMT. Even so, the list of lines was never
sufficient and astronomers continue to ask for more obscure transitions to be
included. A local line catalogue is an excellent back up but it is also
necessary to look up catalogues from remote services. The \almaot\
\citep{2013ASPC..475..373W} works in exactly this way with much success. This
feature will be added in the future using the Simple Line Access Protocol
(SLAP) that will allow for retrieval of line lists from big world-wide atomic
and molecular database such as NIST
\citep{NIST_ASD,2012APS..DMP.D1004K,2004JPCRD..33..177L}, Splatalogue
\citep{2010AAS...21547905R}, and all the databases from the VAMDC consortium
\citep{2011BaltA..20..503K}.

\subsection{Calling user interfaces from pipelines requires forethought}
Interactive user interfaces are very powerful but a number of issues become
apparent when reduction and analysis facilities are directly integrated into an
application where automation is a secondary concern.  At the James Clerk
Maxwell Telescope \splat\ replaced the \specx\ package
\citep[][\ascl{1310.008}]{specx} which had been developed for interactive
command-line use whilst supporting a scripting language to allow for bulk
processing of spectra. With \splat\ the promise of \beanshell\ support was
never able to overcome some key difficulties associated with automation. There
are two scenarios for automation. The first is for an astronomer to record the
steps taken manually in reducing or analyzing a spectrum and then replay that
analysis on many other spectrum. The second scenario is to allow key algorithms
to be made usable in a pipeline environment with the GUI disabled. For the JCMT
heterodyne pipeline
\citep[][\ascl{1310.001}]{2008ASPC..394..565J,JennessACSISDR} all efforts at
algorithmic enhancement were focused on individual \Starlink\ applications such
as \KAPPA\ \citep[][\ascl{1403.022}]{sun95}, because it was unclear how
feasible it would be to add complex algorithms to \splat\ and make them
available through the pipeline. One of the interesting approaches taken by
\gaia\ \citep{2009ASPC..411..575D} was that all analysis algorithms were called
through a messaging interface such that they could be used by the interactive
GUI, from the command-line or from the pipeline. This proved to be a very
powerful approach but was not available in the \splat\ design. Taking care to
separate core algorithms from interactive interfaces is an important design
decision that has long term ramifications for an application.

\subsection{Sub-millimetre units are complicated}
\vounits\ \citep{vounits} standardization is very useful but is not
sufficient for the majority of spectral line data that is encountered
by sub-millimetre astronomers. Sub-millimetre spectra are generally
calibrated in temperature rather than janskys and there are two
competing standards for handling the different calibration factors
that are involved in converting from such definitions as corrected
antenna temperature, main-beam temperature and antenna temperature
\citep{1981ApJ...250..341K,1989LNP...333..351D,2009tra..book.....W}. The
units alone do not tell you which temperature scale is in use and many
efficiency values are required when comparing spectra from different
telescopes. The situation is in dire need of standardization efforts
and we simply had to ignore the issue in \splat.

\subsection{Service selection interface depends on Registry}
The new \ssap\ service selection interface, using more metadata information
from the services, is meant to help the user to choose the specific services
needed and to avoid making unnecessary queries and downloading data which is
not needed. Sometimes too much detail may lead to confusion, especially if the
service information from the registries is often incomplete. If the data is not
there as it should be a detailed interface won't do much. The current interface
will be redesigned in the future, to be probably simpler but more accurate. It
has also been planned to use \regtap\ \citep{regtap} to look for spectral
services, which may provide easier handling.

\subsection{Tracking VO standards is hard}
\splat\ transitioned from a local analysis tool to a VO client during the birth
of the VO and witnessed an explosive growth of standards during that time. One
interesting outcome of this was that funding for \splat\ became scarce as the
VO grew and funds were diverted from classical astronomy software development
efforts. VO standards became increasingly important and it was clear that
support should be added for \ssap\ and related standards although it was a
struggle to prioritize this effort as VO adoption was initially slow and
astronomers were not driving the initial support for these facilities.

\splatvo\ is being updated to add support for standards such as \datalink\
\citep{datalink}, \obscore\ \citep{obstap} and \regtap\ \citep{regtap} as well
as expanded units support with \vounits\ \citep{vounits} required to understand
the metadata from these services, but the VO standards will continue to evolve
as new versions are announced and new standards developed and it will be a
continuing struggle to remain compliant. There is a worry that all the
development effort available to \splatvo\ will be spent simply on maintaining
standards support and this will be at the expense of supporting new analysis
algorithms. This is a very difficult balancing act and we can not expect the
IVOA standards process to stagnate purely to make life easier for applications
developers.

\subsection{The client is not omnipotent}
During the discussions about VO standards we often hear that the protocol
should be very simple, focusing just on allowing users to query the particular
data set and express the location of original files, with all the additional
post-processing of data left as a task for the client.

Unfortunately, when we started using \splatvo\ and a simple \ssap\ server of
stellar spectra from Ond\v{r}ejov HEROS observations, we immediately faced the
limitations of purely client-based processing. Every spectrum of HEROS
\citep{2002PAICz..90....1S} is more than forty five thousand points long
covering about 4500\,$\AA$ (in two channels). Upcoming instruments from
sub-millimetre telescopes such as CCAT \citep{jenness_spie2014} may have a
hundred thousand channels per spectrum and echelle spectra (like those from
UVES \citep{2000SPIE.4008..534D} or HARPS \citep{2000SPIE.4008..582P}) may have
as many as three hundred thousand points. Just downloading thousands of spectra
takes a long time, but zooming on every spectral line to see it in detail and
unzooming takes even longer, while consuming lot of memory.

This practical experience demonstrated the analysis of a stack of hundreds of
complete spectra in \splatvo\ on a common computer, was a very inconvenient and
time consuming task (if possible at all), while usually only a short spectral
range of tens of Angstroms was needed to display each time.  If the server
itself (usually a very powerful computer with a lot of memory) would trim to
the required range, the client need download only a very small data chunk,
which could be quickly displayed.  Instead of spending time zooming to another
line we can in principle perform another query in different spectral range
requesting again only a short piece of another line range.

The need for server-side post-processing, as articulated by our experience with
analysis of thousands of spectra lead to the first
implementation\footnote{\url{http://wiki.ivoa.net/internal/IVOA/InterOpOct2008DAL/stelSSAcutout.pdf}}
of a VO-based post-processing server, based on the \pleinpot\ suite
\citep{2005ASPC..347..385C}, with simple on-the-fly cutouts and relative
rescaling of spectra, later on ported to \dachs\ and supplemented by automatic
continuum
normalization\footnote{\url{http://wiki.ivoa.net/internal/IVOA/InteropMay2012DAL/champaign-getdata.pdf}}.
As \ssap\ does not contain support for post-processing, it was decided
pragmatically to ``overload'' the \ssap\ query parameters representing query
and processing parameters at the same time.

This required to set up another (post-processing) \ssap\ server, where the
\texttt{BAND} limits are used for cutting of the given spectral range
(previously selected) and \texttt{FLUXCALIB} was understood as a command to
divide by the maximum value (\texttt{FLUXCALIB=relative}) or call automatic
continuum normalization (\texttt{FLUXCALIB=normalized}).  This allowed us to
display detailed profiles of hundreds of spectra, including the very long
echelle spectra, study line profiles evolution in long time-series or identify
emission episodes in a long series of Be stars observations just by
overplotting hundreds of on-the-fly normalized cuts of the H$_\alpha$ line.
Although this solution is not consistent with original VO standard motivation,
it served its purpose and resulted in the \texttt{getData} proposal, finally
replaced by \datalink, even though the original \datalink\ was not conceived as
post-processing mechanism (rather it should just point to alternative version
of images or previews etc.).  In any case it is clear that the client is
generally not able to handle large amounts of spectra, due to lack of memory
and processing power. The properly balanced design separation of tasks between
client and server allows for comfortable processing of really Big Data.  A very
nice and clean protocol, which cannot handle real scientific needs, must
finally be made more complex by extensions. Even if it is nice idea, we cannot
put all the processing burden on the client.  This is always the difficult
trade-off between design simplicity and scientific requirements.

\section{Obtaining SPLAT-VO}
\splatvo\ is released in a variety of ways, you can build it using the source
code, obtain it as part of a \Starlink\ JAC release. Alternatively you can
install into your desktop using a standalone IzPack bundled version or run it
up using Java webstart.

The bundled standalone and webstart releases of \splatvo\ are currently
available for the following desktops, Linux (32 and 64 bit), Mac OS X (32 and
64 bit Lion) and Windows Vista and later (32 and 64 bit). Previous versions
also worked on Solaris, Sparc and Intel, and OS X PPC.

The \Starlink\ releases \citep[e.g.,][]{currie_adassxxiii,2013ASPC..475..247B}
are available for Linux (32 and 64 bit) and OS X (Lion and Snow Leopard).

See \ascl{1402.007}\footnote{or \url{http://astro.dur.ac.uk/~pdraper/splat}}
for the main support site where links to these releases can be found.  Releases
of recent GAVO developments are also available and these developments will be
merged into the main \splatvo\ releases.

The source code for \splatvo\ and its associated libraries is open-source and
available on Github at \url{https://github.com/Starlink/starjava}

\section{Conclusions}
\splatvo\ is a very powerful tool for the analysis of spectra allowing the
immediate browsing of world-wide distributed archives as well as detailed
studies of individual objects across the whole electromagnetic spectrum by the
building of SEDs from spectra taken in many different energy bands. Its
capabilities make it an ideal tool for the analysis of astronomical spectra on
the local desktop, facilities that are promoted to full productivity when
combined with resources from the VO environment. It is currently under active
development, promising soon handling of light curves and data cubes. We also
intend to add support for automatic matching of line identifiers to spectra,
simultaneous processing of multiple spectra, automated continuum normalization,
and enhanced support of FITS extensions and metadata.

As a result of combining the effort of GAVO \dachs\ server suite developers and
\splatvo\ developers, \splatvo\ is an ideal reference implementation for
testing new IVOA proposed standards, as required by IVOA rules, for pushing
standards in the recommendation phase.

\section*{Acknowledgements}
The initial work on \splatvo\ was supported by the now closed \Starlink\
Project funded by the Particle Physics and Astronomy Research Council and more
recently by the Joint Astronomy Centre, Hawaii, also funded by the Particle
Physics and Astronomy Research Council and more recently by its successor
organization the Science and Technology Facilities Council.

The current work on \splatvo\ by GAVO is supported by German Federal Ministry
of Education and Research, BMBF grant 05A11VH3 and the Czech development is
supported by grant 13-08195S of Granting Agency of the Czech Republic and by
the Czech project RVO:67985815.

We thank Mark Taylor, who added the \samp\ support to \splatvo\ and created the
VO Registry and \tap\ query libraries used (these are part of StarJava).  We
also thank Markus Demleitner who actively supports the new features in
\splatvo\ \ssap\ and \datalink\ handling by the development of GAVO \dachs\
server publishing suite.  Finally we thank Jan Wouterloot for his work testing
the sub-millimetre features of \splatvo.

Many new features of \splatvo\ are based on everyday experience with analysis
of hot emission line stars spectra observed at Perek 2m Telescope of the
Ond\v{r}ejov observatory of the Astronomical Institute of the Academy of
Sciences, Czech Republic. We greatly acknowledge the help of M.~\v{S}lechta,
responsible for reduction of most of the spectra and students T.~Peterka and
J.~N\'advorn\'\i{}k for the set up and maintenance of the \ssap\ services of
the AI stellar department, with which \splatvo\ was heavily tested in recent
years.

\end{document}